\documentclass[useAMS,usenatbib,fleqn,twocolumn]{aastex7}
\usepackage{multirow}
\usepackage{graphicx}
\usepackage{xcolor,soul}
\usepackage{amssymb,amsmath}
\usepackage{xspace}
\usepackage{tabularx}
\usepackage[utf8]{inputenc}
\usepackage[T1]{fontenc}
\providecommand{\noopsort}[1]{}

\def\sgra{Sgr~A$^*$\xspace}
\def\m87{M87$^*$\xspace}

\usepackage{xcolor}
\usepackage{soul}
\usepackage[normalem]{ulem} 

\begin{document}

\shorttitle{Observing FEEs}
\title{Identifying Observational Signatures of Flux Eruption Events in Supermassive Black Hole Accretion Flows with Machine Learning}

\shortauthors{Ricarte et al.}
\correspondingauthor{Angelo Ricarte}
\email{angelo.ricarte@cfa.harvard.edu}

\author[0000-0001-5287-0452]{Angelo Ricarte}
\affiliation{Center for Astrophysics | Harvard \& Smithsonian, 60 Garden Street, Cambridge, MA 02138, USA}
\affiliation{Black Hole Initiative at Harvard University, 20 Garden Street, Cambridge, MA 02138, USA}
\email{angelo.ricarte@cfa.harvard.edu}

\author[0000-0003-4143-9717]{Erandi Chavez}
\affiliation{Center for Astrophysics | Harvard \& Smithsonian, 60 Garden Street, Cambridge, MA 02138, USA}
\email{erandi.chavez@cfa.harvard.edu}

\author[0009-0005-4236-9636]{Franc O}
\affiliation{Khoury College of Computer Sciences, Northeastern University, Boston, MA 02115, USA}
\affiliation{NSF Institute for Artificial Intelligence \& Fundamental Interactions (IAIFI), Massachusetts Institute of Technology (MIT), Cambridge, MA 02139, USA}
\email{}

\author[0000-0002-8178-8463]{Pavlos Protopapas}
\affiliation{John A. Paulson School of Engineering and Applied Science, Harvard University, Boston, MA 02134, USA}
\email{pavlos@seas.harvard.edu}

\date{\today}

\begin{abstract}
Simulated black hole accretion flows with strong magnetic fields often exhibit ``flux eruption events'' (FEEs), transient and localized expulsions of matter near the event horizon due to magnetic reconnection.  It may now be possible to image them with the Event Horizon Telescope (EHT), a global network of millimeter-wave observatories that images black holes.  Here we use machine learning as an interpretable inference tool to identify observational signatures of FEEs that could be accessible to the EHT.  
First, we train a convolutional neural network to learn task-relevant representations of FEEs in uncorrupted simulated images.  
After using this network to label a larger set of images, we then train interpretable models (random forest and logistic regression) to determine observational signatures. 
We find that during a FEE, images in the millimeter tend toward more diffuse emission, higher linear polarization, and lower total fluxes, but these signatures are weak for most FEEs compared to the usual time variability of these features.  Moreover, the $Q-U$ loop rotation rate decreases during FEEs, contrary to a picture in which FEEs could jointly cause both millimeter $Q-U$ loops and flares.  Our random forest trained on observable summary statistics achieves $\approx$80\% class-weighted accuracy, suggesting that the CNN learns FEE structure not fully mapped onto these traditional summary statistics.  Our results imply that image size and polarization fraction can be used to flag candidate FEEs, but high-resolution, high-dynamic range images will still be important to confirm FEEs and test accretion flows for this phenomenon.
\end{abstract}


\section{Introduction}
\label{sec:introduction}

Using a global array of millimeter-wave telescopes, the Event Horizon Telescope (EHT) collaboration has produced polarized images of two supermassive black holes (SMBHs), \sgra and \m87, each of which exhibits ring morphologies and largely rotationally-symmetric polarization patterns \citep{EHTC+2019a,EHTC+2021a,EHTC+2022a,EHTC+2024b,EHTC+2025}. These images provide new observational constraints to accretion flow models on event horizon scales, which in turn inform models of active galactic nucleus (AGN) fueling and feedback important for galaxy evolution \citep{Kormendy&Ho2013,Heckman&Best2014}.

As the EHT continues to observe both of these targets, ongoing improvements to both the array and image reconstruction algorithms are improving its angular resolution and dynamic range, while opening access to the frequency and time domains.  At the time of writing, multi-wavelength Very Long Baseline Interferometry (VLBI) campaigns of \m87 have occurred, from which visibilities across several months of observation are anticipated.  Known time-domain analyses include pattern speed reconstructions in total intensity \citep{Conroy+2023,Conroy+2025}, complementary signatures in linear polarization \citep{Ricarte+2025b}, and ``jet archaeology'' studies that link jet structures to the event horizon scale processes that produce them \citep{Tsunetoe+2025}.  Yet the time domain remains relatively understudied in simulations, motivating investigation of additional temporal or transient features.

So far, EHT full polarization data, combined with multi-wavelength constraints, have favored models with dynamically important magnetic fields---so-called ``magnetically arrested disk'' (MAD) models over their ``standard and normal evolution'' (SANE) counterparts \citep{EHTC+2021b,EHTC+2022e,EHTC+2024c}.  In MAD models, the poloidal magnetic field threading the horizon saturates, and magnetic tension, pressure, and reconnection transform the dynamics of the inner accretion flow \citep{Bisnovatyi-Kogan&Ruzmaikin1974,Igumenshchev+2003,Narayan+2003}.  Magnetized cavities can form through which inflowing material must snake through non-axisymmetric streams \citep[see e.g., Figure~3 of][]{Wong+2022}, while the strong poloidal field can power very efficient jets powered by the SMBH spin \citep{Blandford&Znajek1977,Tchekhovskoy+2011,Ricarte+2023d,Lowell+2024}.  It is important to continue testing this interpretation, since these aspects of magnetized accretion could have implications for both accretion and feedback efficiencies that may compound on cosmological timescales \citep{Ricarte+2023d,Guo+2023,Cho+2023,Ricarte+2025a,Su+2025,Guo+2025,Cho+2025,Porras-Valverde+2026}.

One characteristic of MAD simulations are ``flux eruption events,'' hereafter FEEs, transient cavities that appear in the innermost accretion flow due to sudden energy injection from magnetic reconnection.  In a typical poloidal field setup, the magnetic field lines anchored to the SMBH must reverse direction at the equator \citep[see e.g., cartoon visualizations of][]{Ricarte+2021b}, leading to an equatorial current sheet that can develop X-points where magnetic reconnection occurs \citep[e.g.,][]{Ripperda+2022}.  FEEs are argued to be the main mechanism for transporting angular momentum in MADs \citep{Chatterjee&Narayan2022,Most&Wang2024,Jacquemin-Ide+2025}.  The same magnetic reconnection that inflates a cavity during a flux eruption event has also been proposed as a mechanism for powering multi-wavelength flares, either by heating surrounding plasma at cavity edges or directly accelerating local electrons due to the transient electric field \citep{Dexter+2020,Porth+2021,Ripperda+2022,Scepi+2022,Zhdankin+2023,Hakobyan+2023,Jia+2023,Najafi-Ziyazi+2024,Vos+2024,Zhou+2025,Solanki+2025,Antonopoulou+2025}.  Only with high-dynamic range resolved imaging on event horizon scales can FEEs be unambiguously detected, and observing them would be a test of both the MAD accretion flow hypothesis as well as the link between flares and magnetic reconnection.  

In this paper, we perform the first comprehensive study of the observational signatures of FEEs.  We train a convolutional neural network (CNN) to detect the presence of FEEs in a library of simulated images at scale.  Then, to determine EHT-accessible observational signatures distinguishing FEE snapshots from quiescent ones, we train a random forest on observable characteristics of these images typically accessible to the EHT.  

We briefly summarize the characteristics of our input GRMHD image library in \autoref{sec:images}.  
In \autoref{sec:neural_network}, we describe how we build and train a CNN to detect FEEs.
\autoref{sec:applications} applies the CNN to the full library and presents the resulting FEE incidence rates and trends in a high-cadence imaging set.
To draw more general conclusions, we train a random forest and use explainability metrics to determine what observable features change during a FEE in \autoref{sec:explainability}. 
We summarize our findings and discuss notable limitations and implications in \autoref{sec:conclusions}.  

\section{Images from GRMHD}
\label{sec:images}

\begin{figure*}
   \centering
   \includegraphics[width=0.9\textwidth]{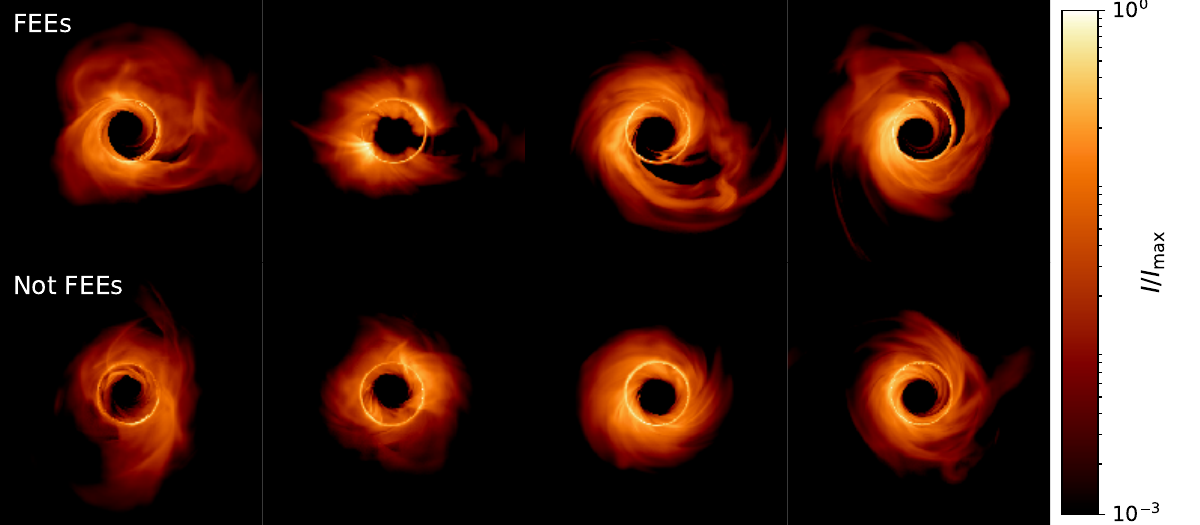}
   \caption{Example FEE and non-FEE snapshots used to train our CNN model, as seen by the model: in logarithmic intensity scaling with 3 decades of dynamic range.  Snapshots are labeled as FEE if there is a visually apparent cavity in the image that extends to the inner shadow (ignoring the photon ring).}
   \label{fig:example_fees}
\end{figure*}

\begin{table*}[thb]
\begin{tabular}{|l|l|l|}
\hline
                        & \textbf{Standard Set}                          & \textbf{High-cadence Set}           \\ \hline
$a_\bullet$             & -0.9, -0.7, -0.5, -0.3, 0, 0.3, 0.5, 0.7, 0.9 & 0, 0.5,* 0.9                \\ 
$R_\mathrm{high}$       & 1, 10, 40, 160                        & 1, 10, 40, 160             \\ 
Magnetic Field State    & MAD, (SANE)                           & MAD                        \\ 
Duration Imaged [$t_g$] & $9 \times 10^4$, ($2 \times 10^4$)    & $10^4$                     \\ 
Snapshots per Model     & 201                                   & 5001                       \\ 
$\nu$ [GHz]             & 228, 345                              & 228, 345                   \\ 
Electron Prescription   & Thermal, Variable $\kappa$            & Thermal, Variable $\kappa$ \\ \hline
\end{tabular}
\caption{Summary of parameters of our GRMHD library.  For each combination of parameters, 201 snapshots have been ray-traced.  A small subset of the ``Standard Set'' has been labeled for use training and validating our network, as described in \autoref{sec:images}.  For the high-cadence set with $a_\bullet=0.5$, only $R_\mathrm{high}=1$ is computed (for analysis in \autoref{sec:qu_loops}).}
\label{tab:grmhd}
\end{table*}

We begin with ray-traced models of \m87 mostly calculated by \citet{Ricarte+2022} using {\sc IPOLE} \citep{Moscibrodzka&Gammie2018} on GRMHD snapshots \citep{Narayan+2022} produced using {\sc KORAL} \citep{Sadowski2013,Sadowski2014}.  We summarize the properties of this library in \autoref{tab:grmhd}.  Relative to \citet{Ricarte+2022}, we have computed additional ``variable $\kappa$'' and higher-cadence images as described below.  Our CNN only operates on total intensity (Stokes $I$) 228 GHz images, although our random forest analysis includes polarization (Stokes $I$, $Q$, $U$, $V$) and the 228 to 345 GHz spectral index, $\alpha \equiv d \log I/ d\log \nu$.

These GRMHD simulations are initialized with tori in hydrostatic equilibrium, differing in dimensionless SMBH spin parameter, $a_\bullet$, for which we sample 9 values, and the emergent magnetic field state, for which we sample both ``MAD'' and ``SANE'' \citep[see][for more details]{Ricarte+2022}.  During ray-tracing, we model the flow as two-temperature, specifying the ion-to-electron temperature ratio as well as the electron distribution function.  The ion-to-electron temperature ratio is set via

\begin{equation}
    \frac{T_i}{T_e} = \frac{1}{1+\beta^2} + R_\mathrm{high}\frac{\beta^2}{1+\beta^2},
\end{equation}

\noindent where the parameter $R_\mathrm{high}$ refers to the asymptotic ion-to-electron temperature ratio at large plasma $\beta = P_\mathrm{gas} / P_\mathrm{mag}$ \citep{Moscibrodzka+2016}.  Given the electron temperature, we then model the electron distribution function as either thermal (Maxwell-J\"uttner) or ``variable $\kappa$'' using radiative transfer coefficients from \citet{Marszewski+2021}.  The latter refers to the relativistic $\kappa$ distribution, which consists of a thermal core and a high-energy power law tail with slope $d\log n / d\log \gamma \propto \kappa-1$ \citep{Vasyliunas1968,Xiao2006}.  We specify $\kappa$ as a function of local plasma $\beta$ and $\sigma = b^2/\rho$ according to the particle-in-cell simulation fitting formulae of \citet{Ball+2018} while conserving internal energy \citep[see also][]{Davelaar+2018,Fromm+2022,EHTC+2022e}.  During ray-tracing, we zero the density in regions with plasma $\sigma > 1$ for the thermal set, or $\sigma > 20$ for the variable $\kappa$ set \citep[to avoid artificial floor material; see e.g.,][]{Wong+2022}.  For each parameter combination, we fit for time-variable fluid rescalings to reproduce an average flux density of 0.5 Jy assuming a mass of $6.2 \times 10^9 \ M_\odot$ and distance of $16.9 \ \mathrm{Mpc}$ as in \citet{Qiu+2023}.  Images are computed with a field of view of 160 $\mu$as and a resolution of 0.5 $\mu$as.

A snapshot is labeled as a FEE \textit{if there is a visually distinct cavity most likely attributable to a reconnection event}.
Images are inspected in Stokes $I$ in logarithmic scale with 3 decades of dynamic range, and are labeled by the lead author by visual inspection. Only the ``Standard Set'' with $R_\mathrm{high}=1$ and thermal electron distributions are used for labeling.  Note that different values of $R_\mathrm{high}$ and the electron prescription use the same underlying GRMHD fluid snapshots.  Example FEE and non-FEE snapshots, scaled as described, are plotted in \autoref{fig:example_fees}. As a cross-check, the second author labeled a randomly selected subset of 50 images, and agreement was obtained for 47/50 of them.

This method of categorization has the advantage of selecting for cavities that are plausibly detectable at the observational frequency of the EHT, but with higher resolution and dynamic range than currently available.  Cavities are generally ubiquitous and long-lived in e.g., the mid-plane density of GRMHD snapshots \citep[see e.g., Figure 4 of][]{Wong+2022}, and this methodology ensures that the labeled cavity produces a recognizable observational signature.  On the other hand, this labeling is subjective and does not provably link magnetic reconnection with the appearance of these cavities.  We expect that a more fluid-centered definition of a flux eruption would not be easy to formulate, and may require higher-cadence GRMHD training data, but could be a future avenue of improvement.

A total of 188 images containing flux eruptions are identified across all 9 spin values and are labeled ``1.''  A total of 218 MAD images without cavities as well as an additional 152 SANE images are labeled ``0.''  Although we do not study SANE models in this paper, we found that including the SANE images improved performance, as they often contain ``grooves'' in total intensity images that are not associated with flux eruptions \citep[e.g., Figure 7 of][]{Ricarte+2022}.  These snapshots are temporally well-separated among the long durations of these simulations and can be treated as independent.  While this is a modest sample size of independent data, the strong geometric features of FEE cavities nevertheless lead to strong performance, as we will describe in the upcoming section.

\section{A Neural Network to Identify FEEs}
\label{sec:neural_network}

We build our inference architecture using ResNet-18, a relatively simple and well-understood CNN \citep{He+2015}.  ResNet-18's skip connections help gradients flow through deeper networks.  While more recent architectures such as Vision Transformers often achieve stronger performance, our goal here is not architectural novelty but a reliable baseline. Moreover, while Vision Transformers may outperform convolutional networks in large-data regimes, the relatively small size of our dataset, and the locality of cavities, favor convolutional architectures with stronger inductive biases.  We initialize the model with the included weights pre-trained on the ImageNet 2012 dataset \citep{Russakovsky+2014}, which contains 1.3 million images categorized into 1000 classes, and fine-tune it for our specific task as described below.

\subsection{Initial Setup}

We make several modifications to ResNet-18 to tailor the architecture to our problem.

\begin{itemize}
    \item ResNet-18 expects RGB images, while our inputs are grayscale (Stokes $I$). We therefore modify the first convolutional layer to accept a single channel, re-initializing weights by averaging the pre-trained weights across the three input channels.

    \item As is standard practice, we replace the final classification head with a single-output logit, $\mathrm{Sigmoid}(\cdot)$ of which is the predicted FEE probability.

    \item ResNet-18 may contain more capacity than needed for this task. For stability, we freeze the parameters of the first block, which are sensitive to generic image features.

    \item Regularization: To reduce overfitting, we apply 30\% dropout in the final fully connected layer.
\end{itemize}

\noindent Additionally, we pre-process our GRMHD images to make them more suitable for this problem and compatible with ResNet-18.

\begin{itemize}
    \item First, we scale the images in logarithmic scale, with three orders-of-magnitude of dynamic range, as in \autoref{fig:example_fees}.  
    \item To match ResNet-18 input expectations, images are first normalized between 0 and 1.  Then, they are each scaled to a mean value of 0.449 and a standard deviation of 0.226 (which are the means of the values expected for the default three channels in ResNet-18). Images are also resized to $224 \times 224$ pixels.
    \item During training, we perform a random resized crop between 70\% and 100\% of the image as an additional augmentation. 
\end{itemize}

Since a FEE is intrinsically a GRMHD fluid-level phenomenon, we reuse the same labels determined for $R_\mathrm{high}=1$ for all 3 other values.  We further augment our dataset by a factor of 8 via 90$^\circ$ rotations and transposes.  Thus, a single GRMHD snapshot is represented 32 times in the dataset.  Including these augmentations, the final dataset used for both training and validation contains 17,856 images.  Nevertheless, we caution that the number of independent fluid snapshots used for training is small (558) and trained only on a single GRMHD library.  While this is appropriate for this labeling task, a larger dataset of independently labeled snapshots would be preferable for deployment as a general tool.  Note that the ``High-cadence Set'' is simply an extension the original simulations; models trained on our ``Standard Set'' do not require extrapolation beyond the training distribution to be applicable to our ``High-cadence Set.''  

\subsection{CNN Training}

We implement five-fold cross-validation, wherein 5 copies of the CNN are separately trained on different 80\% subsets of the labeled data. Each image appears in the validation set exactly once. We ensure that all 32 images derived from the same underlying GRMHD fluid snapshot are assigned to the same fold.

We train by minimizing the weighted binary cross-entropy loss function, 

\begin{equation}
    \mathcal{L} = -\frac{1}{N}\sum_{i=1}^N\left[ y_i \log (p_i) + (1-y_i)\log (1-p_i)\right]w_i,
\end{equation}

\noindent where $y_i$ is an image's true label, $p_i$ is the FEE probability predicted by the model, and $w_i$ is a weight to account for the imbalance between non-FEE and FEE snapshots, equal to non-FEE to FEE ratio for FEEs, and 1 otherwise, so that these two classes have equal weight in the loss function. We use the Adam optimizer with a fixed learning rate of $10^{-4}$ and weight decay of $10^{-4}$.

\begin{figure}
   \centering
   \includegraphics[width=0.5\textwidth]{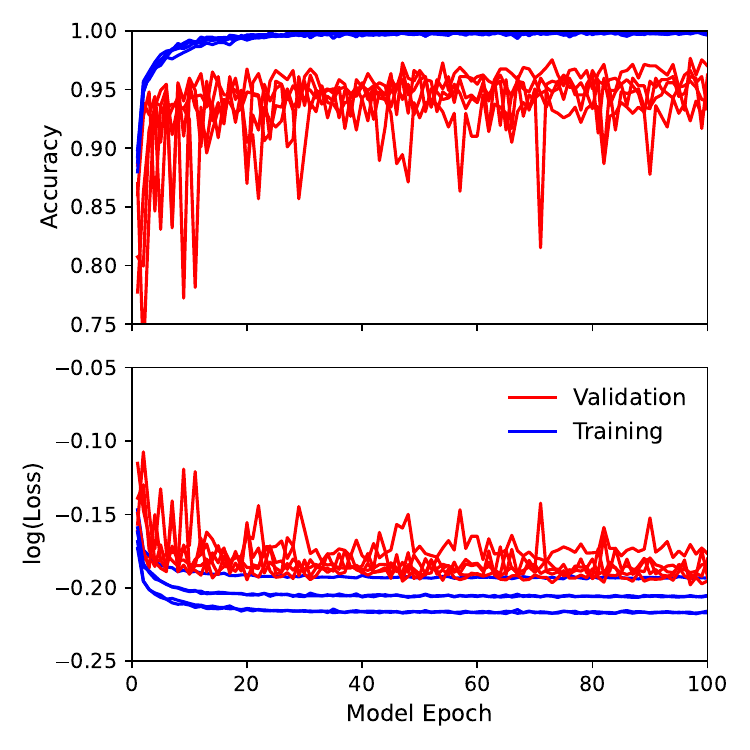}
   \caption{CNN accuracy and loss during training, on both the validation and training sets, as a function of model epoch.  Each track represents one of the five copies of the model.  The rate of improvement rapidly slows around epoch 20.  For science, we keep for each copy the version that attains the minimum loss on the validation dataset.}
   \label{fig:accuracy_loss}
\end{figure}

In \autoref{fig:accuracy_loss}, we plot the accuracy and loss curves exhibited by each of our models during training.  We train in batches of 32 images, but only output results after each epoch of 17,856 images.  Accuracy rapidly approaches 100\% on the training set (blue), but only reaches $\approx$95\% on the validation set (red) before improvement significantly slows.  For each fold, we save the model which achieves the smallest loss on the validation set for use in subsequent sections.  

\begin{figure*}
   \centering
   \includegraphics[width=\textwidth]{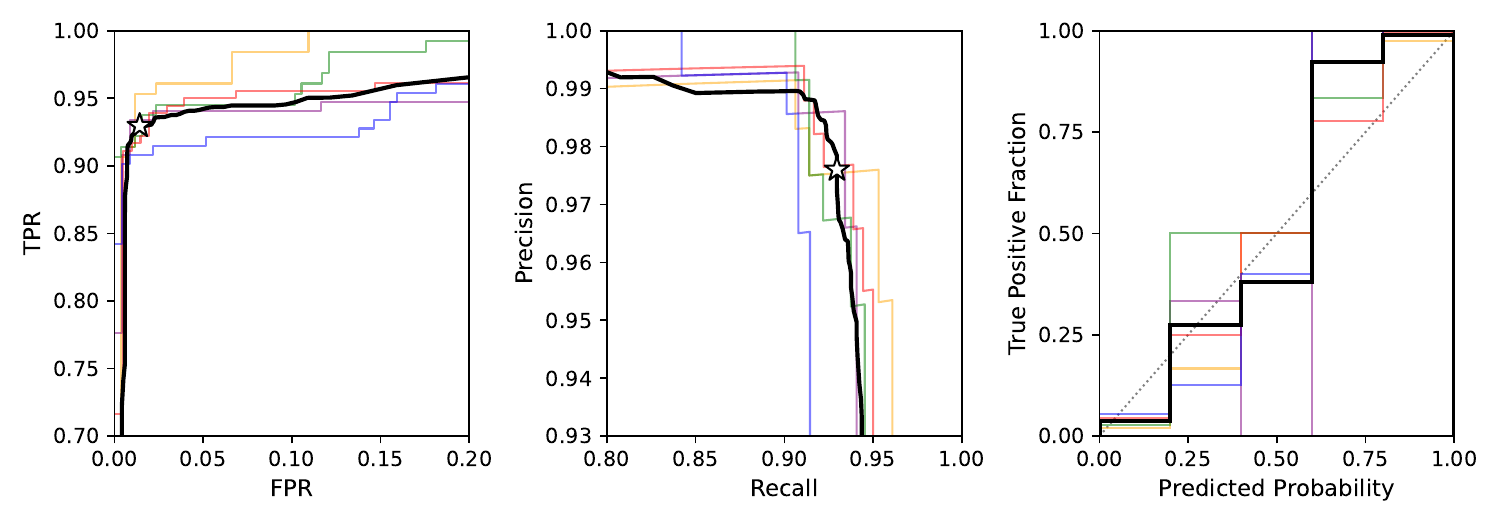}
   \caption{Receiver Operating Characteristic (ROC), Precision-Recall (PR), and calibration curves for each of our 5 CNNs (faint colored lines) as well as the average performance (bold black lines).  Our models are well-calibrated without additional tuning, and summary statistics are provided in \autoref{tab:stats}.}
   \label{fig:diagnostics}
\end{figure*}

To understand the behavior of our CNNs, we plot the receiver operating characteristic (ROC) curve, precision-recall curve (PR), and calibration curve in \autoref{fig:diagnostics} using our validation data.  The ROC curve plots the true positive rate (TPR) against the false positive rate (FPR) as the CNN threshold changes, while similarly the PR curve instead plots precision (true positives over all positives) versus recall (fraction of true positives that are predicted positive).  Each model's curve is plotted with a faint colored line, and the mean behavior is plotted with a bold black line.  We mark with a star the values achieved by our adopted (default) threshold value of 0.5, which naturally achieves a reasonable compromise among these quantities.  Summary statistics are provided in \autoref{tab:stats}, where we further find excellent performance for both ``area under curve (AUC)'' statistics.  The calibration curve in the final column plots the true positive fraction in bins of output probabilities, where a perfectly calibrated model lies exactly along the diagonal.  We do not find significant evidence that post-hoc temperature calibration would be necessary to improve our probabilities \citep[e.g.,][]{Guo+2017}.

Given the subjective nature of our labels, this performance is more than necessary to perform reasonable labeling.  To jointly assess CNN performance and inter-annotator disagreement, the second author performed a blind classification of a randomly selected subset of ``high cadence'' snapshots, which are not used during CNN training. The second annotator identified FEEs substantially more frequently than the CNN trained on the first annotator's labels (29\% versus 16\%), underscoring that the identification of FEEs is inherently subjective and depends on the operational definition adopted. Relative to the second author's labels, the CNN achieved moderately high precision (81\%) but lower recall (45\%). Because the CNN was trained to reproduce the lead-author's annotation scheme, we interpret this weaker performance as reflecting inter-observer variability rather than classifier failure, highlighting the intrinsic ambiguity of FEE identification.  Indeed, the fraction of images assigned a FEE probability between 0.1 and 0.9 is 63\% on images where the second author and CNN disagree compared to 32\% overall.

\begin{table}[]
\centering
\begin{tabular}{|l|l|}
\hline
Statistic & Value \\
\hline
FPR       & 0.014 \\
TPR       & 0.930 \\
Recall    & 0.930 \\
Precision & 0.976 \\
F1        & 0.952 \\
ROC-AUC   & 0.980 \\
PR-AUC    & 0.981 \\
\hline
\end{tabular}
\caption{CNN summary statistics, where performance is averaged among models.}
\label{tab:stats}
\end{table}

\section{Applying the CNN}
\label{sec:applications}

In \autoref{sec:neural_network}, we described how we assembled a set of 17,856 images from GRMHD and adapted a well-known CNN, ResNet-18, to identify the presence of FEEs.  Using 5-fold cross-validation, we produce 5 copies.  Here, we explore results from their initial application.

\subsection{FEEs as a Function of Spin and $R_\mathrm{high}$}

\begin{figure*}
   \centering
   \includegraphics[width=0.9\textwidth]{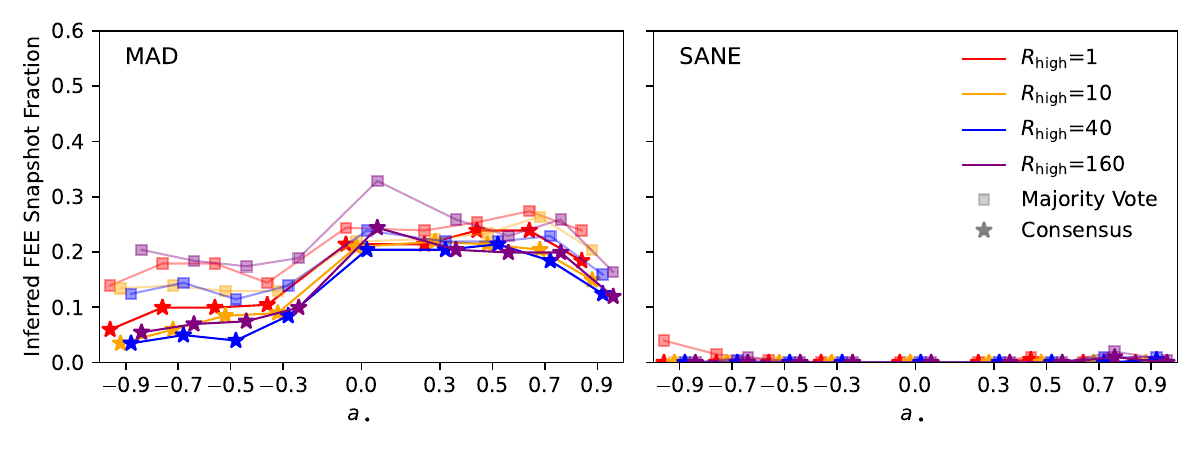}
   \caption{Inferred FEE incidence as a function of $a_\bullet$ and $R_\mathrm{high}$ for both MAD and SANE models.  ``Majority vote'' labeling labels a snapshot a FEE if at least 3/5 CNNs agree that a snapshot is a FEE, while ``consensus'' labeling requires all CNNs to agree.  
   We adopt ``consensus'' labels throughout to obtain a purer sample of FEEs.  The model infers more FEEs for $a_\bullet\geq 0$ than for $a_\bullet < 0$, where there is more ambiguity.  As desired, there is not a significant dependence on $R_\mathrm{high}$.}
   \label{fig:fee_incidence}
\end{figure*}

With 5 CNN models in hand, we apply their predictions to all of the GRMHD snapshots described in \autoref{tab:grmhd}.  In \autoref{fig:fee_incidence}, we plot the FEE snapshot fraction that the model predicts as a function of both $a_\bullet$ (x-axis) and $R_\mathrm{high}$ (colors).  We do not train separate models on the variable $\kappa$ images, instead copying over labels from the thermal images, so only results for thermal models are shown here.  With faint squares, we plot the ``majority vote,'' where an image is labeled a FEE if at least 3/5 CNNs labels it a FEE.  With solid stars, we plot ``consensus,'' requiring all CNNs to agree with a FEE label.  As expected, the ``consensus'' labels are more conservative.  We do not find a strong dependence on $R_\mathrm{high}$, as desired.

For MAD models, there is a much larger gap between ``majority vote'' and ``consensus'' labels for retrograde ($a_\bullet < 0$) models.  This is expected, since retrograde flows tend to be more disordered and ambiguous to classify by eye.  While we do not use them for analysis, SANE images serve as a sanity check on our methodology, as none of them exhibit FEEs.  Both ``majority vote'' and ``consensus'' confidently label almost all SANEs as non-FEEs.  To achieve a more pure sample, we cautiously adopt the more conservative ``consensus'' labels for our analysis. 

We find that the FEE snapshot fraction is remarkably high for MAD models, even adopting conservative ``consensus'' labels.  The overall fraction is $14 \pm 7$\% for this simulation suite and labeling scheme, with larger fractions exhibited by prograde than retrograde models.  The error here corresponds to the standard deviation of this fraction across different values of $a_\bullet$ and $R_\mathrm{high}$. Within this simulation suite, this implies a 50\% chance of observing a FEE after only $\sim$7 independent observations of M87* (under idealized assumptions of independence and detectability).

\subsection{Example: $a_\bullet=0.9$, $R_\mathrm{high}=160$}
\label{sec:example}

\begin{figure*}
   \centering
   \includegraphics[width=\textwidth]{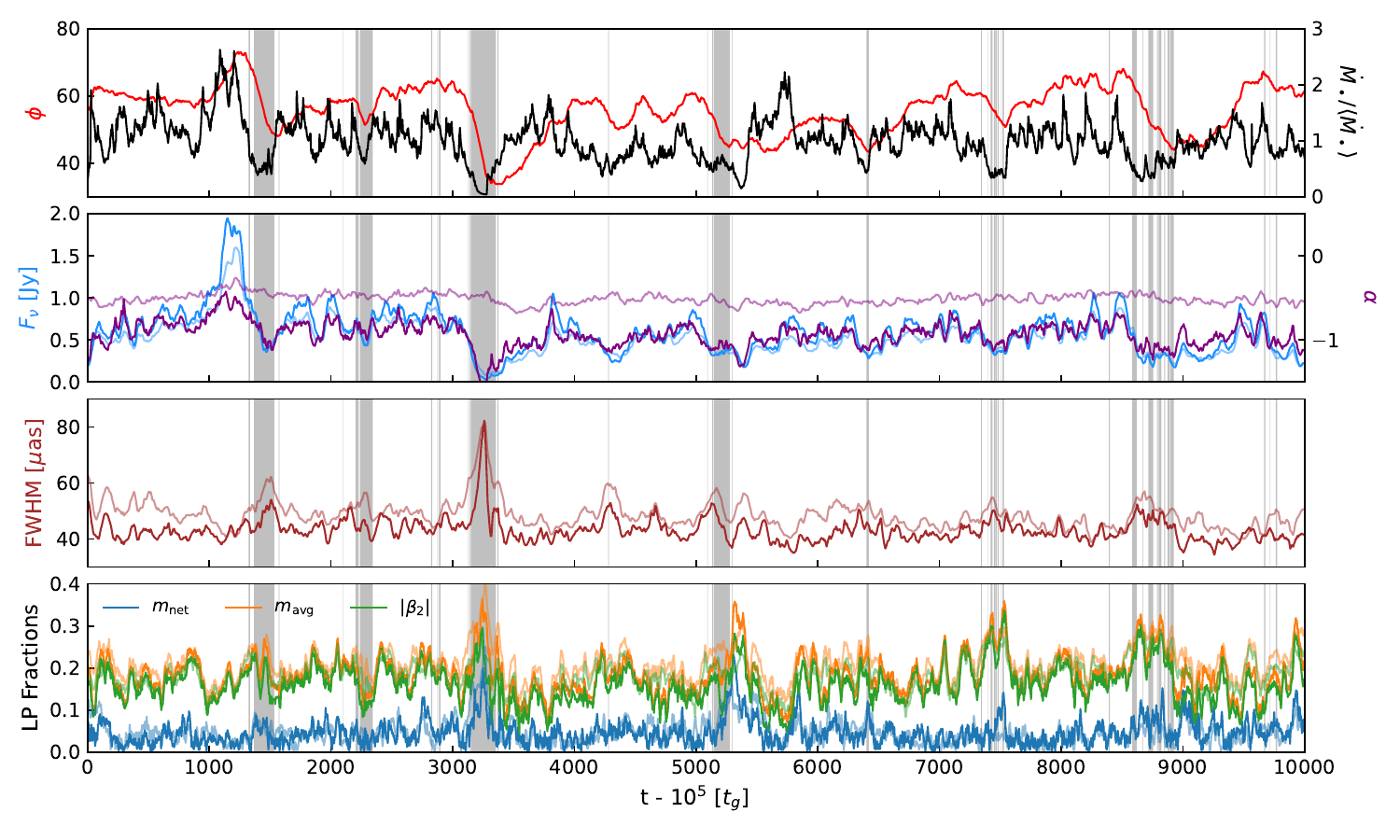}
   \caption{Selected GRMHD and GRRT quantities in our high-cadence data for our $a_\bullet=0.9$, $R_\mathrm{high}=160$ model.  Gray bands are shown where the CNN detects a flux eruption.  Flux eruptions are associated with systematic increases in $\phi$ and decreases $\dot{M}_\bullet$, although this is not directly observable.  The most dramatic flux eruption occurs near the time marked $\sim$3200, when the total flux approaches 0, and the FWHM exceeds 80 $\mu$as.}
   \label{fig:high_cadence}
\end{figure*}

For each snapshot, we compute several different GRMHD and image features to identify physical trends distinguishing FEEs from non-FEEs.  In \autoref{fig:high_cadence}, we plot many of the quantities we explore as a function of time in our high-cadence $a_\bullet=0.9$ $R_\mathrm{high}=160$ models, where thermal models are plotted as solid lines, and variable $\kappa$ models are plotted as fainter lines.  As we will discuss, thermal and variable $\kappa$ models are very similar across most observable features.  The gray bands plot the result of ``consensus'' labeling on our images.

\subsubsection{Fluid Properties}

In the top row, we first plot quantities from the GRMHD snapshots themselves, which are not directly observable, and are shared between thermal and variable $\kappa$ models.  In red, we plot the magnetic flux parameter

\begin{equation}
    \phi(t) = \frac{\sqrt{4\pi}}{2\sqrt{\dot{M}_{\bullet,2000}(t)}} \int_\theta \int_\phi \left|B^r\right|_{r=r_{\rm H}} \;\sqrt{-g}\;\mathrm{d}\phi\;\mathrm{d}\theta,
    \label{eqn:phi_mad}
\end{equation}

\noindent a dimensionless parameter describing the magnetization of the black hole, with $\phi \sim 50$ sometimes used as a definition of the MAD state.  This quantity is evaluated at the event horizon $r_H$ using the accretion rate

\begin{equation}
    \dot{M}_\bullet(t) = -\int_\theta \int_\phi \left[\rho u^r\right]_{r=r_{\rm H}} \;\sqrt{-g}\;\mathrm{d}\phi\;\mathrm{d}\theta, \label{eqn:mdot} \\
\end{equation}

\noindent which we plot in black using a second y-axis.  In \autoref{eqn:phi_mad},  $\dot{M}_{\bullet,2000}(t)$ represents $\dot{M}_\bullet(t)$ convolved with a boxcar filter spanning 2000 $t_g$, constructed so that any rapid time-evolution of $\phi$ is driven by magnetic flux instead of accretion rate. Otherwise, peaks in $\phi$ could correspond to valleys in $\dot{M}_\bullet$ as a definitional consequence rather than an independent correlation.

As expected from the physics of magnetic reconnection, FEEs are associated with decreases in $\phi$.  It is noteworthy that visible cavities appear during much of the downturn in $\phi$, not just its minimum.  Meanwhile, we see that FEEs result in decreases in $\dot{M}_\bullet$, with the most dramatic FEE near $t \approx 3200 \ t_g$ coinciding with a near-cessation of the accretion rate.

\subsubsection{Light Curve}

In the second row, we plot quantities associated with the total intensity light curve that can be measured without resolving the source.  In blue, we plot the 228 GHz flux density, $F_\nu$, while in purple, we plot the spectral index between 228 and 345 GHz, $\alpha$.  For thermal models, we find that $\alpha$ traces $F_\nu$ remarkably well, both of which scale with the optical depth.  Although $F_\nu$ is very similar between thermal and variable $\kappa$ models, we see that $\alpha$ behaves very differently for variable $\kappa$, hovering at a near constant $\alpha \approx -0.5$.  We do not find that FEEs are associated with flares (in the millimeter) using either of our eDF prescriptions, and as we will explore in subsequent sections, FEEs are systematically associated with \textit{decreases} in $F_\nu$.

\subsubsection{Image Size}

In the third row, we plot the full-width at half-maximum (FWHM) of each image (characteristic image size), which requires the source to be spatially resolved.  The FWHM can be estimated without imaging, using the fall-off of visibility amplitudes as a function of baseline length \citep[e.g.,][]{Doeleman+2008,Issaoun2019}.  The inclusion of non-thermal electrons makes variable $\kappa$ models systematically larger than their thermal counterparts, as has been reported by several previous studies \citep[e.g.,][]{Ozel+2000,Mao+2017,Ricarte+2023}.  For both models, we can see that FEEs tend to be associated with increases in image size.  This matches expectations from \autoref{fig:example_fees}, where we can see FEEs push the emitting region to slightly larger radius.

\subsubsection{Image Polarization}

Finally, in the fourth row, we plot several metrics capturing linear polarization.  Consider the Stokes parameters $I$, $Q$, $U$, and $V$.  For each of these Stokes parameters, let us write e.g., $I = \iint I(\rho,\varphi)\rho \mathrm{d}\varphi \mathrm{d}\rho$, where $I$ is the total flux density obtained by summing over the image, and $I(\rho,\varphi)$ refers to the flux density per solid angle subtended by a pixel with solid angle $\rho \mathrm{d}\varphi \mathrm{d}\rho$.  Let the total linear polarization $P = \sqrt{Q^2 + U^2}$ and similarly $P(\rho,\varphi) = \sqrt{Q(\rho,\varphi)^2 + U(\rho,\varphi)^2}$.  We define the ``net'' linear polarization fraction, measurable by a single-dish,

\begin{equation}
    m_\mathrm{net} = \frac{P}{I},
\end{equation}

\noindent and the image ``average'' linear polarization fraction as

\begin{equation}
    m_\mathrm{avg} = \left\langle \frac{P(\rho,\varphi)}{I(\rho,\varphi)} \right\rangle.
\end{equation}

We will later use the equivalents for circular polarization,

\begin{equation}
    v_\mathrm{net} = \frac{V}{I}
\end{equation}

\noindent and 

\begin{equation}
    v_\mathrm{avg} = \left\langle \frac{V(\rho,\varphi)}{I(\rho,\varphi)} \right\rangle.
\end{equation}

We also compute complex $\beta_m$ coefficients \citep{Palumbo+2020} by integrating over the entire image via

\begin{equation}
    \beta_m = \frac{1}{I}\iint P(\rho,\varphi)e^{-im\varphi}\, \rho \mathrm{d}\varphi \mathrm{d}\rho.
\end{equation}

\noindent We are most interested in the $\beta_2$ coefficient, which describes the rotationally-symmetric polarization mode.  Here, $|\beta_2|$ encodes the strength of this mode, while $\angle \beta_2$ encodes the pitch angle.  This well-studied coefficient encodes properties of the magnetic field geometry, velocity field, and space-time \citep[e.g.,][]{Chael+2023,Palumbo2025}.

While $m_\mathrm{net}$ and $v_\mathrm{net}$ correspond to spatially unresolved measurements, $m_\mathrm{avg}$, $v_\mathrm{avg}$ and $\beta_m$ depend on the spatial resolution.  In this work, we compute these quantities only after convolving images with a Gaussian filter with a FWHM of 20 $\mu$as, mimicking the present resolution of the EHT at 230 GHz.

During a FEE, we anticipate larger polarization fractions since shock heating and internal energy injection from magnetic reconnection should locally increase the temperature and reduce Faraday depolarization if this heating is enough to move a significant fraction of electrons into the ultrarelativistic regime \citep{Jones&Odell1977}.  Indeed we find that each of these metrics tends to increase, although the effect is subtle.  This is likely because the aforementioned heating is local, may not necessarily act on the Faraday depolarizing plasma, and only matters if it results significant changes in the suppression factor $\approx\ln(\Theta_e)/\Theta_e^2$, where $\Theta_e = k_bT_e/m_ec^2$.  A decrease in the optical depth may also play a role.

\section{Making Observational Predictions}
\label{sec:explainability}

\subsection{Identifying Predictive Features Using a Random Forest}
\label{sec:random_forest}

Since the CNN described in \autoref{sec:neural_network} operates on uncorrupted Stokes $I$ images, it requires higher-fidelity information than currently accessible in realistic images.  To identify observationally-accessible predictors of a FEE, we train random forests to predict a FEE using the features described in \autoref{sec:example}, all of which are accessible to the current EHT.  We segment the data such that one random forest is trained on the thermal standard set, and another on the variable $\kappa$ standard set.  Our overall strategy is to progressively move from more complete but opaque image representations to less complete but interpretable observable summaries.

We use \texttt{sklearn.ensemble.RandomForestClassifier} with 100 estimators and a maximum tree depth of 5.  A random forest will obtain 100\% accuracy on its data if allowed to reach arbitrary depth, but this is over-fitting.  We determine our tree depth by first testing the random forest's accuracy against a validation set consisting of 20\% of the data, selecting the largest value for which the class-weighted accuracy improves (before over-fitting occurs).  Then, we fit the complete dataset after determining the optimal tree depth.  The random forest is given a subset of the features listed in \autoref{tab:logistics_thermal}, first pre-processed via Z-score normalization to aid in interpretability by isolating within-model variations in time:

\begin{equation}
    x \mapsto z(x) = \frac{x-\langle x \rangle}{\sigma(x)}.
\end{equation}

When doing so, the mean $\langle x \rangle$ and standard deviation $\sigma(x)$ is computed individually for each combination of $a_\bullet$ and $R_\mathrm{high}$.  In addition, although these quantities are not directly measurable, we pass the random forest both $a_\bullet$ and $R_\mathrm{high}$.  This allows the random forest to adapt its predictions as a function of these parameters and reveal potential dependencies on $a_\bullet$ and $R_\mathrm{high}$.

We include two previously undefined features here, the optical depth $\tau_I$ and the Faraday (rotation) depth $\tau_F$, obtained by first integrating the absorptivity and Faraday rotation coefficients along each geodesic, then performing an intensity-weighted average across the whole image.  In each pixel,

\begin{equation}
    \tau_I(\rho,\varphi) = \int \alpha_I(\lambda) \, \mathrm{d}\lambda
\end{equation}

\noindent and 

\begin{equation}
    \tau_F(\rho,\varphi) = \int \rho_V(\lambda) \, \mathrm{d}\lambda,
\end{equation}

\noindent where $\alpha_I(\lambda)$ and $\rho_V(\lambda)$ are the absorptivity and Faraday rotation coefficients at the position along the geodesic corresponding to affine parameter $\lambda$.  A single scalar for each image is calculated via an intensity-weighted average:

\begin{equation}
    \tau_I = \frac{1}{I}\iint \tau_I(\rho,\varphi)I(\rho,\varphi) \,  \rho \mathrm{d}\varphi \mathrm{d}\rho
\end{equation}

\noindent and

\begin{equation}
    \tau_F = \frac{1}{I}\iint \tau_F(\rho,\varphi)I(\rho,\varphi) \,  \rho \mathrm{d}\varphi \mathrm{d}\rho.
\end{equation}

\noindent For additional discussion of these definitions, see e.g., \citet[][]{EHTC+2021b}.

To interpret the behavior of each random forest, we compute the SHapley Additive exPlanations (SHAP) values of each feature of each image \citep{Lundberg&Lee2017}.\footnote{We use \texttt{shap.TreeExplainer} with the ``interventional'' setting corresponding to \citet{Janzing+2020}.}  In a team game, SHAP is a way to estimate each teammate's contribution to the final score.  During the prediction step (after the forest has already been trained), it calculates how much the outcome changes when the teammate joins every possible combination of other teammates, then averages those changes.  In our problem, each teammate is a feature, and the outcome is FEE (1) or non-FEE (0).  To interpret a given prediction ``$\mathrm{pred}$'' that the random forest has made on a given GRMHD snapshot, the SHAP of a given feature $i$ is calculated 

\begin{equation}
    \mathrm{SHAP}_i(\mathrm{pred}) = \frac{1}{n!}\sum_R \left[\mathrm{pred}(P_i^R \cup \{i\}) - \mathrm{pred}(P_i^R) \right],
    \label{eqn:shap}
\end{equation}

\noindent where the sum includes all possible feature orderings $R$ of which there are $n!$ combinations, and $P_i^R$ is the set of all features preceding $i$ in that ordering $R$.\footnote{SHAP is calculated without retraining the random forest, so to calculate $\mathrm{pred}(P_i^R)$, at each branch in a decision tree the algorithm marginalizes over the conditional distribution of the missing features.}  This can be interpreted as how much (magnitude) and in which direction (sign) each feature ``pushes'' the outcome.  The SHAP of a group of features is simply the sum of the SHAPs of each feature.

\begin{figure*}
   \centering
   \includegraphics[width=0.9\textwidth]{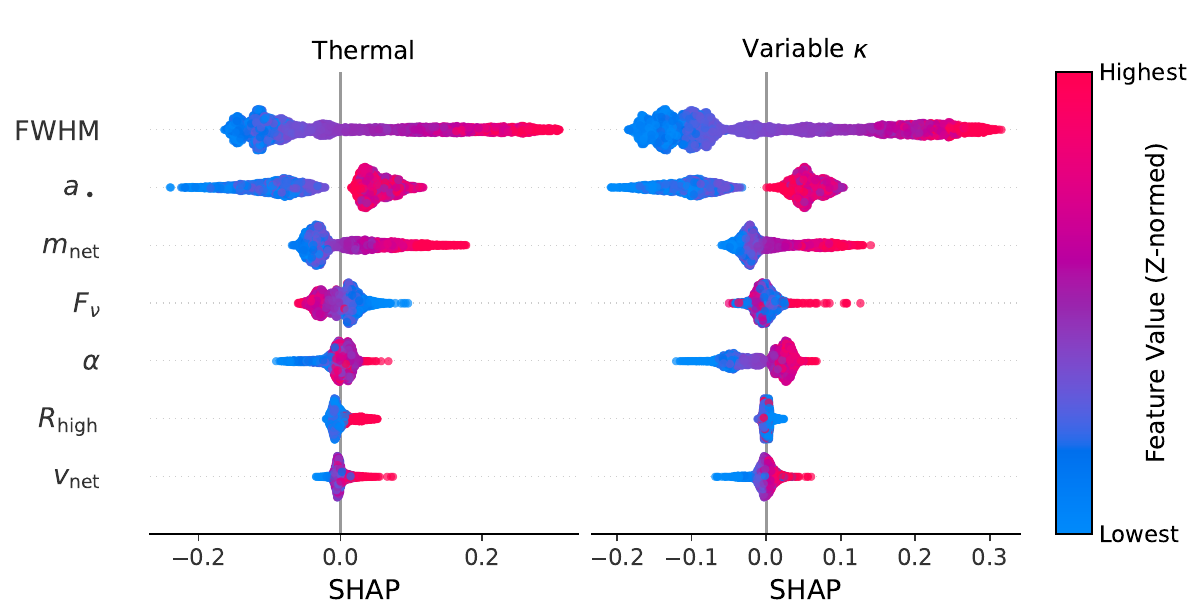}
   \caption{``Bee swarm'' plot depicting the distributions of SHAP values (\autoref{eqn:shap}) for each feature among the images used to train the random forest.  These encode how much and in which direction the prediction is ``pushed'' based on the feature value.}
   \label{fig:shap}
\end{figure*}

In \autoref{fig:shap}, we visualize the distribution of SHAPs computed on the random forests trained on the set of features $\{\mathrm{FWHM}, a_\bullet, m_\mathrm{net}, F_\nu, \alpha, R_\mathrm{high}, v_\mathrm{net} \}$, which we found to achieve class-weighted accuracies of 78\% and 79\% for thermal and variable $\kappa$ models respectively.  We explore other variations in \autoref{sec:random_forest_variants}, but highlight here a compact set that produces near-maximal performance.  Features are placed in descending order of $\langle |\mathrm{SHAP}|\rangle$ across all images.  We find that $\mathrm{FWHM}$ is the strongest predictor, and this feature even outperforms $\dot{M}$ and $\phi$ in a more inclusive variant explored in \autoref{sec:random_forest_variants}.  The bimodal SHAP distribution as a function of $a_\bullet$ reveals that retrograde and prograde images behave significantly differently, likely reflecting the decreased likelihood of FEEs for retrogrades revealed by \autoref{fig:fee_incidence}.  In \autoref{sec:random_forest_variants}, we find that restricting the sample to non-retrograde systems does not change the discussion here.

For subsequent features, trends in the prediction as a function of SHAP mostly follow physical trends that we later validate with single-parameter logistic fits.  As expected, $m_\mathrm{net}$ increases during a FEE.  However, when interpreting SHAPs of non-primary features, it is important to bear in mind that the random forest does not use each feature in a vacuum.  The magnitude of a feature's SHAP may be small, even if it is actually a good predictor of a FEE, if a correlated feature is an even better predictor.  For example, while we verify that $\alpha$ is systematically lower in FEE compared to non-FEE snapshots in the thermal set, the SHAPs reveal mixed behavior, since SHAP measures the \textit{additional} information this feature provides in the context of the other available information, and it is highly correlated with $F_\nu$ (\autoref{fig:high_cadence}).  

\begin{figure}
   \centering
   \includegraphics[width=0.45\textwidth]{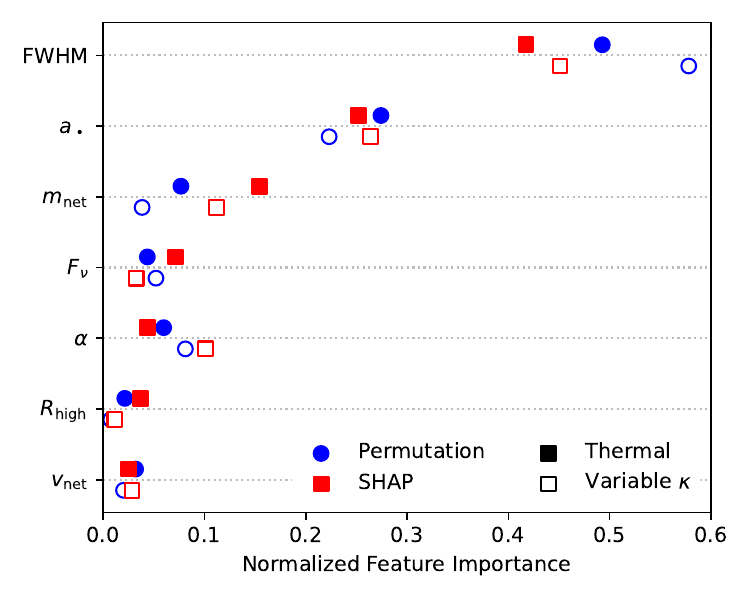}
   \caption{Feature importance derived from our random forest model predicting the incidence of a flux eruption.  Feature importance derived from both SHAP and permutation are provided, which yield very similar results.  FWHM, $m_\mathrm{net}$, and $F_\nu$ are the most impactful observable predictors.}
   \label{fig:feature_importance}
\end{figure}

In \autoref{fig:feature_importance}, we use two different methods to estimate feature importance.  Shown as squares, one method is simply to compute $\langle |\mathrm{SHAP}|\rangle$.  Alternatively, shown as circles, we also measure the decrease in class-weighted accuracy when one feature's vector is randomly shuffled.  To compare these methods with each other, we normalize each such that the sum of feature importances is 1 for each method.  Thermal models are plotted with filled markers, and variable $\kappa$ models are plotted with open markers.  We find that the permutation and SHAP metrics behave very similarly and results echo our discussion of \autoref{fig:shap}.

\subsection{Verifying Trends with Logistic Regression}

As previously discussed, SHAP values of correlated features can be non-trivial to interpret.  To clearly verify physical trends, we also fit single-parameter logistic functions,

\begin{equation}
    f(x) = \frac{1}{1+e^{-k(x-x_0)}},
    \label{eqn:logistic}
\end{equation}

\noindent to the Z-score normalized features that we use for our random forests. Results are presented in \autoref{tab:logistics_thermal} for thermal models and \autoref{tab:logistics_variablekappa} for variable $\kappa$ models.  Error bars are estimated by bootstrapping the sample.  In most cases, $x_0 \approx 0$, implying that the midpoint occurs at the center of the distribution, consistent with what we'd expect for Z-score normalized quantities.  Of greater interest are the sign of $k$ and the class-weighted accuracy achieved with just a single logistic. 

The best performing single-parameter logistic operates on FWHM, achieving a modest class-weighted accuracy of 70\%, while each other logistic performs much worse.  If $k>0$, the feature value typically increases during a FEE.  As previously discussed, this is the case for FWHM and the linear polarization metrics including $m_\mathrm{net}$ and $m_\mathrm{avg}$.  Evidently, the $|\beta_1|$ and $|\beta_2|$ are poorer predictors than these simpler polarization metrics.  In the fluid, we verify that both $\dot{M}$ and $\phi$ systematically decrease during a FEE.  We verify that $F_\nu$ on average decreases during FEEs for both thermal and variable $\kappa$ models.  We are surprised by the weak performance of $\alpha$, which should be sensitive to the temperature and optical depth \citep{Ricarte+2023}, in both thermal and variable $\kappa$ models.  In the variable-$\kappa$ models $k$ is almost consistent with zero, reflecting the near-flat $\alpha \approx -0.5$ visible in Fig.~\ref{fig:high_cadence}.  Indeed, we find that the optical and Faraday depths, $\tau_I$ and $\tau_F$, also both decrease on average during a FEE, but themselves also would serve as poor predictors if observable.  For both sets of models, the circular polarization fractions $v_\mathrm{net}$ and $v_\mathrm{avg}$ are the least predictive features.

In summary, we find that increases in FWHM and the linear polarization fraction (measured via $m_\mathrm{net}$ and $m_\mathrm{avg}$) are the best predictors of a FEE.  However, our random forests trained on summary statistics only achieve up to 80\% accuracy.  While these summary statistics could be used to provide FEE candidates, ultimately there is no substitute for imaging at high resolution and high dynamic range to confirm the presence of a FEE.

\begin{table*}[]
\centering
\begin{tabular}{|lcccl|}
\hline
Feature      & $k$      & $x_0$     & Class-weighted Accuracy & Observability \\
\hline
FWHM    & $0.956 \pm 0.044$  & $-0.292 \pm 0.022$ & $0.697 \pm 0.008$ & VLBI   \\
$\dot{M}$   & $-0.625 \pm 0.041$ & $-0.137 \pm 0.017$ & $0.618 \pm 0.008$ & Not Directly Measurable   \\
$m_\mathrm{avg}$          & $0.564 \pm 0.036$  & $-0.114 \pm 0.015$ & $0.614 \pm 0.007$ & VLBI Polarization   \\
$m_\mathrm{net}$          & $0.485 \pm 0.033$  & $-0.090 \pm 0.013$ & $0.605 \pm 0.007$ & Single-dish Polarization   \\
$F_\nu$           & $-0.241 \pm 0.047$ & $-0.024 \pm 0.010$ & $0.559 \pm 0.009$ & Single-dish   \\
$|\beta_1|$        & $0.245 \pm 0.036$  & $-0.024 \pm 0.008$ & $0.547 \pm 0.010$ & VLBI Polarization   \\
$\phi$           & $-0.194 \pm 0.037$ & $-0.013 \pm 0.005$ & $0.546 \pm 0.010$ & Not Directly Measurable   \\
$\tau_I$          & $-0.178 \pm 0.038$ & $-0.014 \pm 0.006$ & $0.542 \pm 0.008$ & Not Directly Measurable   \\
$|\beta_2|$        & $0.138 \pm 0.037$  & $-0.008 \pm 0.005$ & $0.533 \pm 0.009$ & VLBI Polarization   \\
$\tau_F$          & $-0.162 \pm 0.032$ & $-0.010 \pm 0.004$ & $0.530 \pm 0.008$ & Not Directly Measurable  \\
$\alpha$ & $-0.143 \pm 0.040$ & $-0.008 \pm 0.004$ & $0.529 \pm 0.009$ & Single-dish   \\
$v_\mathrm{net}$          & $0.134 \pm 0.039$  & $-0.007 \pm 0.004$ & $0.527 \pm 0.010$ & Single-dish Polarization   \\
$v_\mathrm{avg}$          & $0.123 \pm 0.036$  & $-0.006 \pm 0.003$ & $0.521 \pm 0.008$ & VLBI Polarization   \\
\hline
\end{tabular}
\caption{Results of single-parameter logistic fits to thermal models to isolate physical trends.  \autoref{eqn:logistic} defines $k$ and $x_0$, and the sign of $k$ encodes whether a feature increases or decreases during a FEE.}
\label{tab:logistics_thermal}
\end{table*}

\begin{table*}[]
\centering
\begin{tabular}{|lcccl|}
\hline
Feature      & $k$      & $x_0$     & Class-weighted Accuracy & Observability \\
\hline
FWHM    & $1.063 \pm 0.046$  & $-0.306 \pm 0.024$ & $0.717 \pm 0.008$ & VLBI   \\
$\dot{M}$   & $-0.602 \pm 0.042$ & $-0.126 \pm 0.018$ & $0.614 \pm 0.008$ & Not Directly Measurable   \\
$m_\mathrm{avg}$          & $0.672 \pm 0.037$  & $-0.155 \pm 0.017$ & $0.630 \pm 0.009$ & VLBI Polarization   \\
$m_\mathrm{net}$          & $0.456 \pm 0.031$  & $-0.079 \pm 0.012$ & $0.600 \pm 0.008$ & Single-dish Polarization   \\
$F_\nu$           & $-0.189 \pm 0.039$ & $-0.013 \pm 0.006$ & $0.550 \pm 0.009$ & Single-dish   \\
$|\beta_1|$        & $0.238 \pm 0.031$  & $-0.022 \pm 0.006$ & $0.545 \pm 0.008$ & VLBI Polarization   \\
$\phi$           & $-0.197 \pm 0.030$ & $-0.014 \pm 0.005$ & $0.546 \pm 0.008$ & Not Directly Measurable   \\
$\tau_I$          & $-0.273 \pm 0.045$ & $-0.030 \pm 0.010$ & $0.575 \pm 0.008$ & Not Directly Measurable   \\
$|\beta_2|$        & $0.286 \pm 0.036$  & $-0.032 \pm 0.008$ & $0.567 \pm 0.009$ & VLBI Polarization   \\
$\tau_F$          & $-0.260 \pm 0.037$ & $-0.024 \pm 0.008$ & $0.540 \pm 0.008$ & Not Directly Measurable  \\
$\alpha$ & $0.074 \pm 0.039$  & $-0.002 \pm 0.002$ & $0.524 \pm 0.010$ & Single-dish   \\
$v_\mathrm{net}$          & $0.167 \pm 0.034$  & $-0.011 \pm 0.004$ & $0.538 \pm 0.009$ & Single-dish Polarization   \\
$v_\mathrm{avg}$          & $0.060 \pm 0.031$  & $-0.002 \pm 0.002$ & $0.512 \pm 0.009$ & VLBI Polarization   \\
\hline
\end{tabular}
\caption{As \autoref{tab:logistics_thermal}, but for variable $\kappa$ models.  There are no notable differences, and we confirm that $F_\nu$ decreases on average during FEEs even in the variable $\kappa$ model.}
\label{tab:logistics_variablekappa}
\end{table*}

\subsection{Flux Eruptions and $Q-U$ Loops}
\label{sec:qu_loops}

\begin{figure}
   \centering
   \includegraphics[width=0.45\textwidth]{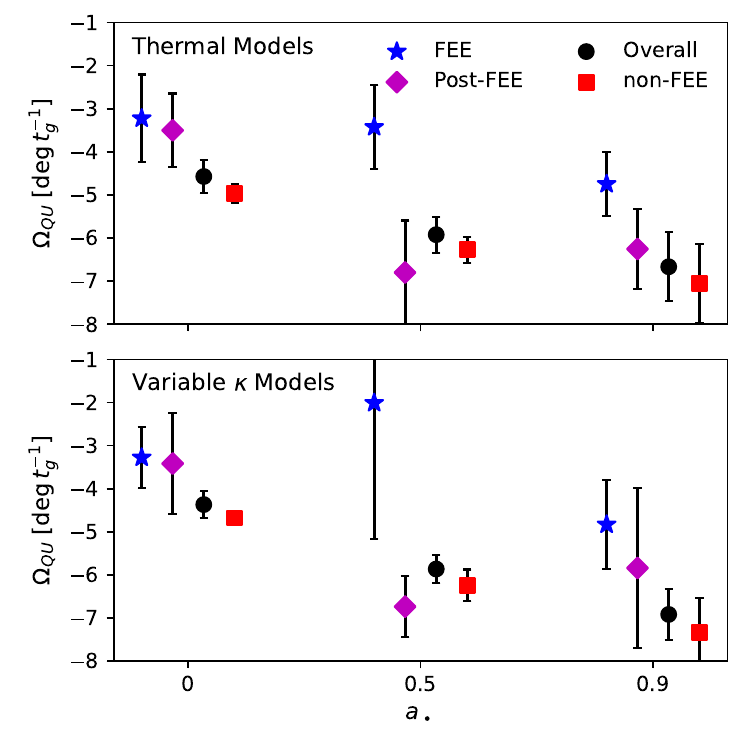}
\caption{Rotation rate in the $Q-U$ plane, $\Omega_{QU}$, as a function of $a_\bullet$ for our high-cadence $R_\mathrm{high}=1$ models, where we compare overall, FEE, post-FEE, and non-FEE periods.  We find that the magnitude of $\Omega_{QU}$ on average {\it decreases} during a FEE, as it disrupts the rotational motion of the accretion flow.  This is consistent with the joint observational and theoretical study of \citet{Ricarte+2025b}, which reported that polarization loops occur continuously, even in the absence of flares.}
   \label{fig:qu}
\end{figure}

When following the polarized flux density of \sgra as a function of time, visually evident loops in the $Q-U$ plane can be observed in the millimeter \citep{Marrone+2006,Wielgus2022QU}, IR \citep{Michail+2026}, and NIR \citep{GRAVITY+2018,Gravity+2023}.  A proposed explanation for this phenomenon is the formation of a transient orbiting polarized hotspot of emission \citep{Broderick&Loeb2006,Gravity+2020b,Vos2022,Yfantis+2024b,Michail+2026}, which could be potentially associated with the magnetic reconnection that triggers a FEE \citep{Najafi-Ziyazi+2024,Grigorian&Dexter2024}.  

To investigate this hypothesis, we compute the $Q-U$ rotation rate, $\Omega_{QU}$, as defined by \citep{Ricarte+2025b}, during FEE and non-FEE periods.  This method quantifies $\Omega_{QU}$ using the differential geometry of planar curves.  In short, the method averages the signed curvature of the curve parametrized by $Q(t)$ and $U(t)$, only setting a ``speed limit'' to eliminate cusps that occur due to finite time resolution.  Using this method, \citet{Ricarte+2025b} found evidence of clockwise motion in the plane during all 4 days for which high-cadence polarized light curves were acquired of \sgra in the \citet{Wielgus2022QU} 2017 dataset.  Because the Gaussian curvature depends only on the local first and second derivatives of these curves, we can use this method to compare $\Omega_{QU}$ among FEE and non-FEE snapshots.  

Results are shown in \autoref{fig:qu} for our high-cadence thermal and non-thermal models with $R_\mathrm{high}=1$.  The overall $\Omega_{QU}$, shown in black, ranges from around $-4 \ \mathrm{deg\,t_g^{-1}}$ to $-7 \ \mathrm{deg\,t_g^{-1}}$, roughly consistent with the values obtained \citet{Ricarte+2025b} using a different set of GRMHD simulations scaled to \sgra.  The negative sign encodes clockwise motion in the $Q-U$ plane, and consequently, the sky.  Error bars are crudely estimated by splitting the time series into 3 different windows each of length 5000 $t_g$ and taking the standard deviation.  We find a clear systematic \textit{decrease} in the magnitude of $\Omega_{QU}$ among FEE snapshots.  This is consistent with the findings of both \citet{Najafi-Ziyazi+2024} and \citet{Ricarte+2025b}, who each report continuous $Q-U$ looping activity in ray-traced GRMHD simulations.  We also do not find evidence that $|\Omega_{QU}|$ increases post-FEE, defined here as any snapshots 50 $t_g$ following a FEE that lasts at least $25 \ t_g$ (based on the definitions in the upcoming section). To explain our findings, we propose that FEEs result in an injection of radial momentum that temporarily disturbs the regular rotation of the inner accretion flow.

The result is nearly identical in our variable $\kappa$ models compared to our thermal models, although we caution that our electron prescriptions, both for their temperature and distribution function, remain highly uncertain.  Even the variable $\kappa$ distribution implemented here, while derived from PIC simulations, is unlikely to fully capture the electron behavior during transient events such as FEEs.  On the other hand, there are no indications that improving our non-thermal eDF prescriptions would promote increased $\Omega_{QU}$ during FEEs.

\subsection{FEE Durations and Gaps}
\label{sec:gaps}

For both observational prospects and comparing with analytic expectations, it is of interest to estimate the duration and gaps between FEEs.  Analytic arguments relating the inflow timescale with the magnetic flux on the horizon predict that major FEEs should recur approximately every $\sim$$10^{3-4}\ t_g$ \citep{Dexter+2020,Jacquemin-Ide+2025}.  This would correspond to timescales of at least years for M87*, or at least weeks for Sgr A*.  Visual examination of \autoref{fig:high_cadence} confirms that long-lived FEE periods are separated by timescales of $\sim$$10^3 \ t_g$.  We attempt to quantify this in our high-cadence set.

We consider the ``consensus'' labels calculated for our $R_\mathrm{high}=1$ high-cadence set (consisting of time series of 0s and 1s).  For defining when a continuous FEE starts and ends, the main challenge is the noise of our CNN predictor, which can produce spurious events with false positives, or split events with false negatives.  To help mitigate this, we first convolve these time series with a boxcar filter of length 7 (corresponding to $14 \ t_g$, where the cadence is $2 \ t_g$), then round the results back to 0 or 1.  We define a continuous FEE to be any contiguous sequence of 1s.  We merge any continuous FEEs separated by less than $14 \ t_g$.  Finally, we reject any continuous FEEs with duration less than $14 \ t_g$.  With a greater volume of labeled data, training models with temporal context could be a future area of improvement.

\begin{figure*}
   \centering
   \includegraphics[width=\textwidth]{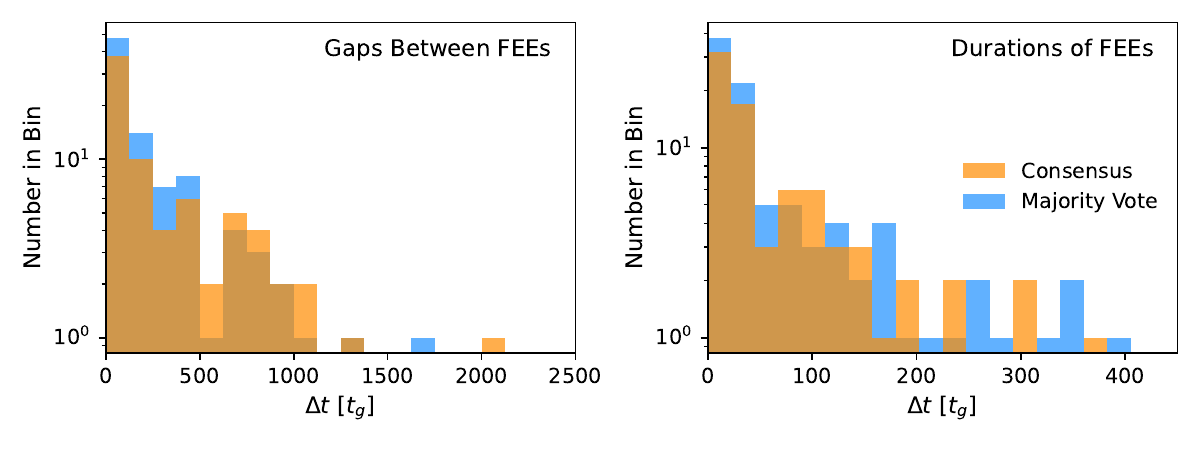}
   \caption{Distributions of delay times between (top) and durations of (bottom) FEEs among our high-cadence thermal $R_\mathrm{high}=1$ models.  While gaps may last up to $\sim$$1000 \ t_g$, delays of only $\sim$$100 \ t_g$ are more common.  Cavities usually persist only a few $10$s of $t_g$, but the longest may last $100$s of $t_g$.}
   \label{fig:gaps}
\end{figure*}

After this processing we have 78 continuous FEEs with durations of at least $14 \ t_g$ by construction.  The median duration is $30 \ t_g$, and the median gap between FEEs is $104 \ t_g$.  The majority (66 of the 78) have a duration of at least $28 \ t_g$ following our processing.  We present a histogram of continuous FEE gaps and durations from this analysis in \autoref{fig:gaps}.  For comparison, we present results using both ``consensus'' and ``majority vote'' labels, and we do not report significant qualitative differences.  This may be because our high-cadence data do not include retrogrades, where a greater disagreement between these labeling schemes was found in \autoref{fig:fee_incidence}.  For either methodology, counts are dominated by both short gaps and short durations.  The longest continuous FEEs can last for 100s of $t_g$, but most are shorter than $30 \ t_g$.  

\section{Discussion and Conclusions}
\label{sec:conclusions}

In this work, we investigate observational signatures of FEEs in MAD GRMHD simulations.  These are transient cavities associated with magnetic reconnection near the horizon, and have been proposed as explanations for both flares and $Q-U$ loops.  To study them, we train a CNN to identify them at scale in ray-traced GRMHD simulations, then perform a random forest and feature importance analysis to make observational predictions for the EHT.  The CNN is used to identify FEEs at scale, while the random forest identifies combinations of observables predictive of FEEs.  Our results can be summarized as follows:

\begin{itemize}
    \item Our CNN finds that $14\% \pm 7\%$ of MAD snapshots contain visually identifiable FEEs, with a larger fraction inferred for $a_\bullet \gtrsim 0$ ($19\% \pm 3\%$) than for $a_\bullet < 0$ ($7\%\pm2\%$). However, the absolute rate should be interpreted in the context of the subjectivity of these labels.

    \item A random forest trained on summary statistics that are accessible to EHT today is able to infer FEEs with a class-weighted accuracy of $80\%$.  The most important observable features distinguishing FEEs and non-FEEs are the FWHM and linear polarization fraction, both of which systematically increase during FEEs.
    
    \item We find that the 228 GHz flux density and spectral index are both poor predictors of a FEE, and that the average behavior is a decrease in the flux density.  This is true not only with thermal eDFs, but also using a ``variable $\kappa$'' non-thermal eDF prescription motivated by PIC simulations.  It is expected that the flux density at higher frequencies (such as NIR and X-ray) should respond more strongly to temperature changes that could occur during FEEs and could be an avenue of future improvement.
    
    \item We find that FEEs are systematically associated with \textit{decreases} to the $Q-U$ rotation rate, likely due to FEEs disrupting the orbital motion of the inflowing material.  We do not find evidence that the large scale FEEs studied in this paper are associated with $Q-U$ loops.

    \item While analytic arguments predict gaps of $\sim$$10^{3-4} \ t_g$ between major FEEs, we find a median gap of only 104 $t_g$ between events, which have a median duration of 30 $t_g$.  This is due to more frequent minor events clustering during active periods.

\end{itemize}

Our results depend on a visual definition of FEEs and a single underlying GRMHD library.  Extending to other simulations and labeling schemes could be an interesting future direction.  We do not test for out-of-distribution generalization of our classifier and expect that training with additional labeled GRMHD-derived images could be necessary before deployment as a general tool.  All quantities in \S\ref{sec:applications}--\S\ref{sec:gaps} are computed using CNN-derived labels.  Consequently, any residual classification errors will propagate into the inferred FEE incidence rates, durations, gap statistics, and derived correlations. Given the high cross-validated classifier performance, we expect these effects to introduce systematic uncertainties at the few-percent level, subdominant to uncertainty from inter-annotator disagreement, but they should nevertheless be borne in mind when interpreting the quantitative values reported in this work.

In addition, our results specifically apply to FEEs as defined by large-scale, image-resolved cavities, and do not exclude the possibility that smaller-scale reconnection events could contribute to observable variability.  Our models are not high enough resolution to resolve plasmoids that can form in a reconnecting current sheet, which have been hypothesized to manifest hotspots that would be absent in our GRMHD simulations \citep{Ripperda+2020,Ripperda+2022}.  Similarly, because the simulations are evolved in ideal GRMHD, magnetic reconnection in these models is mediated by numerical dissipation rather than explicit kinetic or resistive microphysics.

In ray-tracing, modeling of the electron distribution function adds additional uncertainties. The \citet{Moscibrodzka+2016} prescription is meant to approximate the average behavior of the ion-to-electron temperature ratio in two-temperature simulations, but it is unknown how applicable this approximation is during FEEs.  Magnetic reconnection is also expected to accelerate a non-thermal high energy tail of electrons, and while we attempt to model this in post-processing \citep{Ball+2018}, this is again a steady-state prescription based on simulations that do not capture the full dynamics of our system.  

With major events separated by $\sim$$10^3 \ t_g$ and minor ones occurring more frequently, it is possible that FEEs could have already happened during EHT observations, but that it currently lacks the sensitivity to detect them.  When attempting to identify FEEs using summary statistics accessible to the EHT, a class-weighted accuracy of only 80\% is achieved.  This implies image size and linear polarization fraction could be used to flag FEE candidates, but summary statistics alone are insufficient to identify FEEs robustly without high-resolution, high-dynamic range imaging.  These capabilities could be enabled by improvements to the array both on the ground \citep[ngEHT;][]{Doeleman+2019} and into space \citep[BHEX;][]{Johnson+2024}.

\begin{acknowledgements}
We thank Sara Issaoun and Cora Prather for comments that improved the manuscript.  We thank Ramesh Narayan for sharing his GRMHD snapshots.  
This research is funded in part by the Gordon and Betty Moore Foundation, Grant GBMF-12987.
We acknowledge financial support from the National Science Foundation (AST-2307887).
This work was supported by the Black Hole Initiative, which is funded by grants from the John Templeton Foundation (Grant \#62286) and the Gordon and Betty Moore Foundation (Grant GBMF-8273)---although the opinions expressed in this work are those of the author(s) and do not necessarily reflect the views of these Foundations. 
\software{
PyTorch \citep{paszke2019},
eht-imaging \citep{chael2018},
Astropy \citep{astropy2013,astropy2018,astropy2022},
NumPy \citep{harris2020},
Matplotlib \citep{hunter2007}
}
\end{acknowledgements}

\section{Data Availability Statement}

Most of the analysis scripts used to generate the figures and tables in this work are available at https://github.com/ARRicarte/fee\_classifier.git. Public software packages used in the analysis are listed in the Software section.

\appendix

\section{Random Forest Variants}
\label{sec:random_forest_variants}

\begin{figure*}
   \centering
   \begin{tabular}{ll}
   \includegraphics[width=0.6\textwidth]{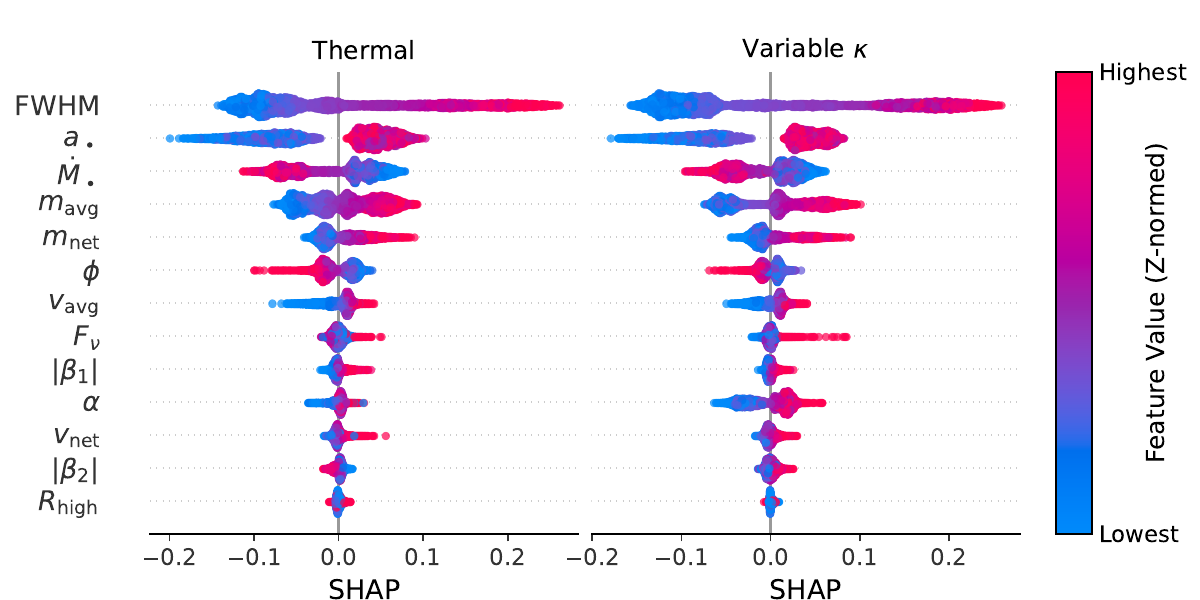} &
   \includegraphics[width=0.35\textwidth]{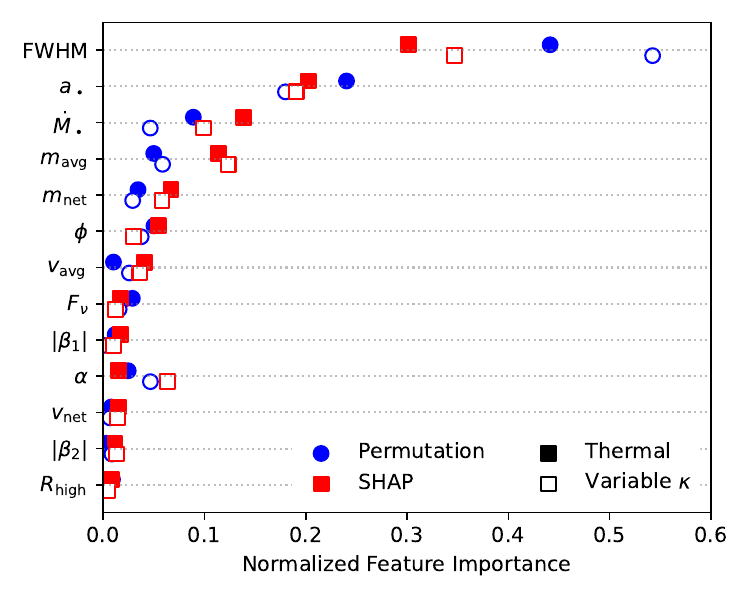} 
   \end{tabular}
   \caption{As Figures \ref{fig:shap} and \ref{fig:feature_importance}, but including a more extensive list of features.  Accuracy does not improve, and the same trends are preserved, implying these extra features do not provide additional useful information.}
   \label{fig:variant_everything}
\end{figure*}

\begin{figure*}
   \centering
   \begin{tabular}{ll}
   \includegraphics[width=0.6\textwidth]{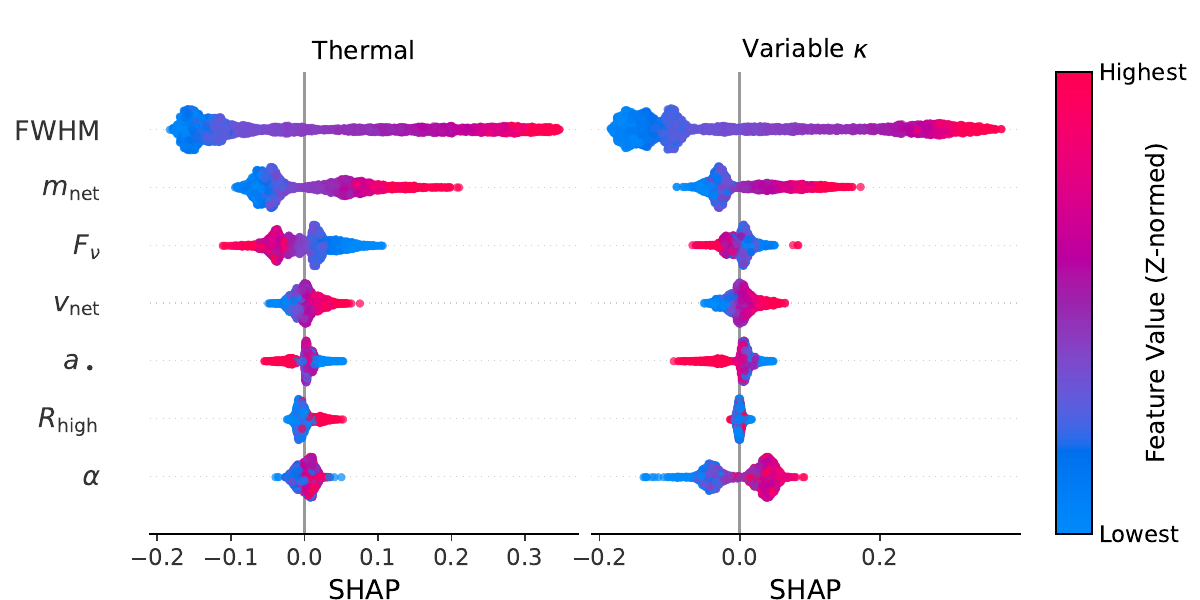} &
   \includegraphics[width=0.35\textwidth]{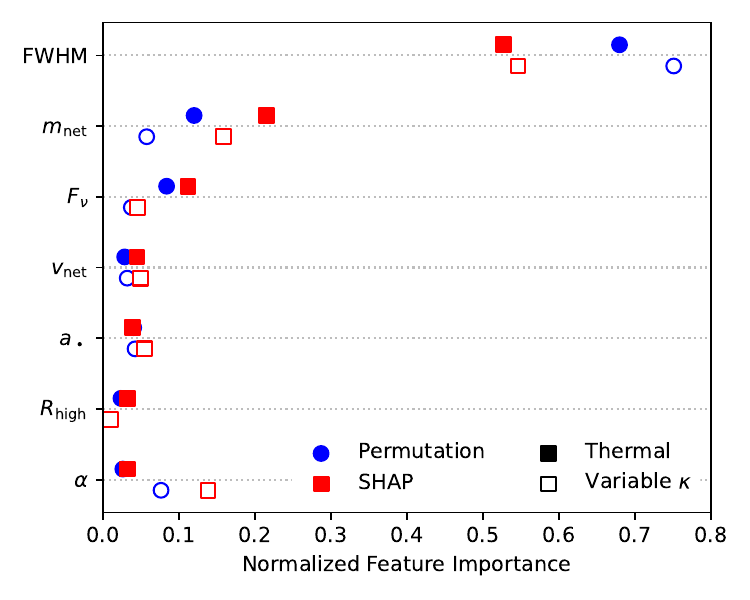} 
   \end{tabular}
   \caption{As Figures \ref{fig:shap} and \ref{fig:feature_importance}, but excluding retrograde models.  This confirms that the prograde vs. retrograde division dominates the feature importance of $a_\bullet$ in \autoref{fig:shap}.}
   \label{fig:variant_noretrogrades}
\end{figure*}

In \autoref{sec:random_forest}, we hand-picked a set of summary statistics to which we trained a random forest to infer the existence of a FEE.  We explore a more extensive variant in \autoref{fig:variant_everything} and a variant excluding retrogrades in \autoref{fig:variant_noretrogrades}.

\autoref{fig:variant_everything} achieves a class-weighted accuracy of 80\% for both the thermal and variable $\kappa$ models, which is a negligible performance increase compared to the simpler model presented in \autoref{fig:shap}.  This is likely due to strong colinearities among the additional features added with those originally included in \autoref{fig:shap}.  It is remarkable that FWHM is a better indicator than either pure-fluid quantities $\dot{M}$ or $\phi$, both of which decrease during a FEE.  Otherwise, features perform as expected based on our logistic regression analysis.

Both \autoref{fig:shap} and \autoref{fig:variant_everything} reveal that $a_\bullet$ is one of the most important parameters for the random forest, showing a clear divergence between progrades and retrogrades.  In contrast, the random forest does not consider $R_\mathrm{high}$ an important feature, consistent with \autoref{fig:fee_incidence}.  Since the SHAP values indicate an important difference between progrades and retrogrades, we test a random forest trained on only $a_\bullet \gtrsim 0$ models in \autoref{fig:variant_noretrogrades}.  Very similar to the model presented in \autoref{fig:shap}, this variant achieves class-weighted accuracies of 79\% and 81\% for thermal and variable $\kappa$ respectively.  We find that $a_\bullet$ plummets in importance, verifying that the prograde-retrograde split dominated this signal.  Otherwise, trends are broadly consistent with \autoref{fig:shap}.

\bibliography{ms}

\end{document}